\documentclass[aps,prl,superscriptaddress,twocolumn]{revtex4-1}
\usepackage{amsmath,amssymb,mathtools}
\usepackage{color}
\usepackage{graphicx}
\usepackage{subfigure}
\usepackage{hyperref}
\usepackage{mathrsfs}
\usepackage{amsthm}
\usepackage{float} %设置图片浮动位置的宏包
\usepackage{ulem}
\def \be {\begin{equation}}
\def \ee {\end{equation}}
\def \ba {\begin{array}}
\def \ea {\end{array}}
\def \bea{\begin{eqnarray}}
\def \eea{\end{eqnarray}}
\def \nn {\nonumber}

\def \td {\tilde}

\begin{document}

%\title{\textbf{Distinguishably of canonical ensemble, microcanonical ensemble, and primary excited states in 2D CFT}}
\title{\textbf{Entanglement of Purification and Projective Measurement in CFT}}
\author{Wu-zhong Guo}
\email{wzguo@cts.nthu.edu.tw}
\affiliation{Physics Division, National Center for Theoretical Sciences, National Tsing Hua University,\\No.\ 101, Sec.\ 2, Kuang Fu Road, Hsinchu 30013, Taiwan}
%\author{
%Wu-zhong Guo$^{1}$\footnote{wzguo@cts.nthu.edu.tw}~
%}
%\date{}

\vspace{-10mm}
%\begin{center}
%{\it
%$^{1}$Physics Division, National Center for Theoretical Sciences, National Tsing Hua University,\\
%No.\ 101, Sec.\ 2, Kuang Fu Road, Hsinchu 30013, Taiwan\\\vspace{1mm}
%}
%\vspace{10mm}
%\end{center}

\begin{abstract}
We investigate entanglement of purification in conformal field theory. By using Reeh-Schlieder theorem, we construct a set of the purification states for $\rho_{AB}$, where $\rho_{AB}$ is reduced density matrix for subregion $AB$ of a global state $\rho$. The set can be approximated by acting all the unitary observables,located in the complement of subregion $AB$, on the global state $\rho$, as long as the global state $\rho$ is \text{cyclic} for every local algebra, e.g., the vacuum state. Combining with the gravity explanation of unitary operations in the context of the so-called surface/state correspondence, we prove the holographic EoP formula. We also explore the projective measurement with the conformal basis in conformal field theory and its relation to the minimization procedure of EoP. Interestingly, though the projective measurement is not a unitary operator, the difference in some limits between holographic EoP and the entanglement entropy after a suitable projective measurement is a constant $\frac{c}{3}\log 2$ up to some contributions from boundary. This suggests the states after projective measurements may approximately be taken as the purification state corresponding to the minimal value of the procedure.
\end{abstract}
\maketitle
\newpage

\section{Introduction}
Recent studies on the gravity dual of some information-theoretical quantities have provided us more insights on the nature of gravity and AdS/CFT correspondence \cite{Maldacena:1997re}. Quantum entanglement in the field theory has a mysterious relation to the definition of geometry in the bulk. In AdS/CFT, the entanglement entropy is given by the area of a minimal surface in AdS\cite{Ryu:2006bv}\cite{Hubeny:2007xt}.\\
The entanglement in quantum field theory (QFT) has a deep relation with the structure and symmetry of the theory. In the framework of algebraic QFT, the constructions of the theory are by observables rather than the states\cite{Haag}\cite{Streater}. Along with this aspect the celebrated Reeh-Schlieder theorem give a strong constraint on the local properties of QFT. In fact this theorem characterizes the strong entanglement in vacuum state between different subregions .\\
In this paper we will use Reeh-Schlieder theorem to investigate a quantity called entanglement of purification (EoP), which is another good entanglement measurement even for mixed state\cite{Terhal}. Similar as entanglement entropy this quantity is also proposed to have geometric interpretation in the context of AdS/CFT. The holographic EoP is proposed in \cite{Takayanagi:2017knl}\cite{Nguyen:2017yqw}.\\
EoP is a quantity to characterize the correlation between different subsystems $A$ and $B$ for a given state $\rho$. For a subsystem $A$ the reduced density matrix $\rho_{A}$  is defined as $\rho_{A}=tr_{\bar A} \rho$, where $\bar A$ is the complement of $A$. The entanglement entropy $S_A$ is given by the von Neumann entropy
\be S(\rho_A):= -tr \rho_A\log \rho_A. \ee
%The mutual information $I(A:B)=S(\rho_A)+S(\rho_B)-S(\rho_{AB})$ can be taken as a physical quantity to characterize the correlation between $A$ and $B$.\\
The entanglement of purification is defined as
\be\label{definitionPurification}
E_{P}(\rho_{AB})= \min\limits_{\rho_{AB}=tr_{\td A \td B} |\psi\rangle \langle \psi|}  S(\rho_{A \td A}),
\ee
where the states $|\psi\rangle$ are called purifications of $\rho_{AB}$ by introducing $\td A$ and $\td B$, and  $\rho_{A \td A}:=tr_{B\td B}|\psi\rangle \langle \psi|$.  The minimization is taken over all the possible purifications $|\psi\rangle$.\\
 The holographic EoP  is given by the area of the minimal cross of entanglement wedge, denoted by $\Sigma_{AB}$,
\be\label{Heop}
E_W(\rho_{AB})= \frac{\text{min}\{\text{area}(\Sigma_{AB})\}}{4G},
\ee
where $G$ is the Newton constant. In this construction the entanglement wedge is the region surrounded by $AB$ and the minimal surface homologous to them, which is expected to be dual to reduced density matrix $\rho_{AB}$\cite{Czech:2012bh}-\cite{Dong:2016eik}.\\
The calculation of EoP in QFT is very hard \cite{Caputa:2018xuf}, for some simple models we may rely on numerical calculations\cite{Hauschild}\cite{Bhattacharyya:2018sbw}. In this paper we find a set of the purification states $|\psi\rangle$ by using Reeh-Schlieder theorem. The set is constructed by unitary transformations in the complement of $AB$. Using this result and combining with the surface/state correspondence \cite{Miyaji:2015yva}\cite{Miyaji:2015fia}, we prove the conjecture of holographic EoP (\ref{Heop}). In the end we also point out the possible relation between projective measurement \cite{Rajabpour:2015uqa}\cite{Rajabpour:2015xkj} and the minimization procedure of EoP.

\section{Reeh-Schlieder Theorem and Purification}
\begin{figure}[H]
\centering
\includegraphics[trim = 0mm 160mm 0mm 0mm, clip=true,width=10.0cm]{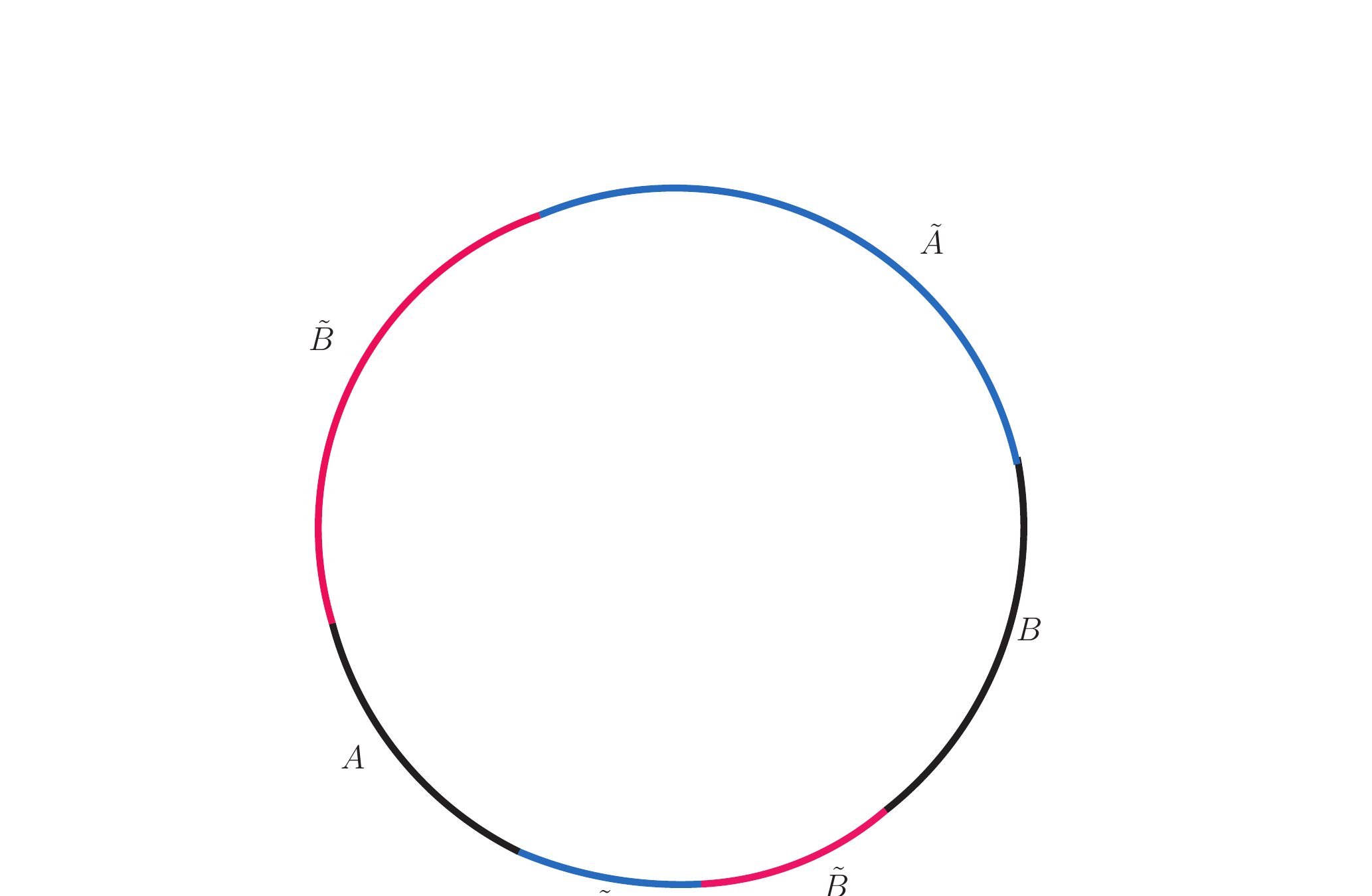}
\caption{A possible division of $\overline{AB}$ by  $\td A$ (red) and $\td B$ (blue). }
\label{f1}
\end{figure}

The minimization procedure (\ref{definitionPurification}) makes the calculation of EoP become a very difficult task in QFT since, in principle, we have to deal with infinite states. Actually there is no method to systematically construct the states $|\psi\rangle$.\\
But this problem will be much easier if the state of the entire system $\rho$ is \textit{cyclic}. To explain the what is meant by a cyclic state we need some basic elements of algebraic QFT\cite{Haag}, see also \cite{Kay}. In the framework of algebraic QFT any  open region $O$  can be associated with a von Neumann algebra of local observables, denoted by $\mathscr{R}(O)$. For $O$ being the entire space region, we have a global algebra $\mathscr{U}$. We denote the Hilbert space of QFT by $\mathcal{H}$. A state $|\Psi\rangle$ is said to be cyclic for $\mathscr{R}(O)$ with respect to the Hilbert space $\mathcal{H}$, if the set $\mathcal{H}_{O}:=\{ \mathcal{O} |\Psi\rangle, \mathcal{O}\in \mathscr{R}(O)\}$ is dense in $\mathcal{H}$. In other words, any state $|\Psi'\rangle\in \mathcal{H}$ can be approximated by the elements in set $\mathcal{H}_{O}$  as closely as we like. For example, the vacuum state $|0\rangle$ is a cyclic state for the global algebra $\mathscr{U}$. But the Reeh-Schlieder theorem gives a much stronger conclusion than that, it shows the vacuum state $|0\rangle$ is also a cyclic state for \textit{every} local algebra $\mathscr{R}$. More precisely, \\
~\\
 \textbf{Reeh-Schlieder Theorem:} \\
  Suppose $O$ to be any bounded open region, then the vacuum state $|0\rangle$ is cyclic for $\mathscr{R}$(O).\\
  ~\\
 One may refer to \cite{Streater} for the proof of this theorem, see also a more modern treatment\cite{Witten:2018lha}. 
 The reason for vacuum state being cyclic for local algebra  is that different regions are highly entangled in vacuum state.
 Now we come back to our discussion of purification and its relation to Reeh-Schlieder theorem. $\rho_{AB}$ is the reduced density matrix of the cyclic state $\rho=|0\rangle\langle 0|$. Firsly, we could show the set of the purification states $|\psi\rangle$ of $\rho_{AB}$ can be approximated by the elements in
\be
\mathcal{H}_{\overline{AB}}:= \{ \mathcal{O}_{\overline{AB}}|0\rangle, \mathcal{O}_{\overline{AB}}\in \mathscr{R}(\overline {AB})\},
\ee
where $\mathcal{O}_{\overline{AB}}$ is the operator located in the region $\overline{AB}$. The Reeh-Schlieder theorem guarantees the set $\mathcal{H}_{\overline{AB}}$ is dense in $\mathcal{H}$. This means  any $|\psi\rangle$ can be approximated by a state $\mathcal{O}_{\overline{AB}}(\psi)|0\rangle$ in $\mathcal{H}_{\overline{AB}}$. We simply write it as\cite{Explain1}
\be
|\psi\rangle= \mathcal{O}_{\overline{AB}}(\psi) |0\rangle.
\ee
We may choose the auxiliary parts $\td A \td B$ as $\overline{AB}$. Fig.\ref{f1} shows one of the possible division of $\overline{AB}$. To satisfy the constraint of purification $tr_{\td A\td B}|\psi\rangle \langle \psi|= \rho_{AB} $, we could further show the operator $\mathcal{O}_{\overline{AB}}(\psi)$ must be  unitary, i.e.,
$\mathcal{O}_{\overline{AB}}(\psi)=\mathcal{U}_{\overline{AB}}(\psi)$ with $\mathcal{U}^\dagger_{\overline{AB}}(\psi)\mathcal{U}_{\overline{AB}}(\psi)=\mathcal{U}_{\overline{AB}}(\psi)\mathcal{U}^\dagger_{\overline{AB}}(\psi)=\bf{1}$.\\
The constraint $tr_{\td A\td B}|\psi\rangle \langle \psi|= \rho_{AB}$ is equal to
\bea
tr_{AB} (O_{AB}tr_{\td A\td B}|\psi\rangle \langle \psi|)=tr_{AB} (O_{AB}\rho_{AB}),
\eea
for arbitrary operator $O_{AB}\in \mathscr{R}(AB)$.
This leads to
\bea\label{RS1}
&&\langle0| (\mathcal{O}_{\overline{AB}}\mathcal{O}^\dagger_{\overline{AB}}-{\bf{1}})\mathcal{O}_{AB}|0\rangle=0,\nn \\
 &&\langle0| (\mathcal{O}^\dagger_{\overline{AB}}\mathcal{O}_{\overline{AB}}-{\mathbf{1}})\mathcal{O}_{AB}|0\rangle=0,
\eea

where we have used the cyclic property of trace and the microcausality condition for local algebra, i.e., $[\mathcal{O}(x),\mathcal{O}(y)]=0$ when $x$ and $y$ are spacelike separated\cite{Haag}. Since (\ref{RS1}) is true for any operator $\mathcal{O}_{AB}$, using the Reeh-Schlieder theorem again, there should exist an operator $\mathcal{O}_{AB}$ such that $\mathcal{O}_{AB}|0\rangle=(\mathcal{O}_{\overline{AB}}\mathcal{O}^\dagger_{\overline{AB}}-{\bf{1}})|0\rangle$ and $\mathcal{O}_{AB}|0\rangle=(\mathcal{O}^\dagger_{\overline{AB}}\mathcal{O}_{\overline{AB}}-{\bf{1}})|0\rangle$. Therefore, by using (\ref{RS1}), the norm of states $(\mathcal{O}_{\overline{AB}}\mathcal{O}^\dagger_{\overline{AB}}-{\bf{1}})|0\rangle$ and $(\mathcal{O}^\dagger_{\overline{AB}}\mathcal{O}_{\overline{AB}}-{\bf{1}})|0\rangle$ are vanishing\cite{Explain2}. Finally, we have
\be
\mathcal{O}^\dagger_{\overline{AB}}\mathcal{O}_{\overline{AB}}=\mathcal{O}_{\overline{AB}}\mathcal{O}^\dagger_{\overline{AB}}={\bf{1}}.
\ee
Now we arrive at our main result in this section.\\
~\\
 { \textbf{Corollary 1:}}\\
 The set of the purifications of reduced density matrix $\rho_{AB}$ can be approximated by the Hilbert space $\mathcal{H}_{\psi}$ constructed by acting unitary local operators $\mathcal{U}_{\overline{AB}}$ on the vacuum, i.e.,
\be\label{purset}
\mathcal{H}_{\psi}= \{ \mathcal{U}_{\overline{AB}}|0\rangle, \quad  \text{unitrary}\quad  \mathcal{U}_{\overline{AB}}\in \mathscr{R}(\overline{AB})\}.
\ee
\section{Surface/State Correspondence and proof of Holographic EoP}
\begin{figure}[H]
\centering
\includegraphics[trim = 0mm 140mm -20mm 0mm, clip=true,width=10.0cm]{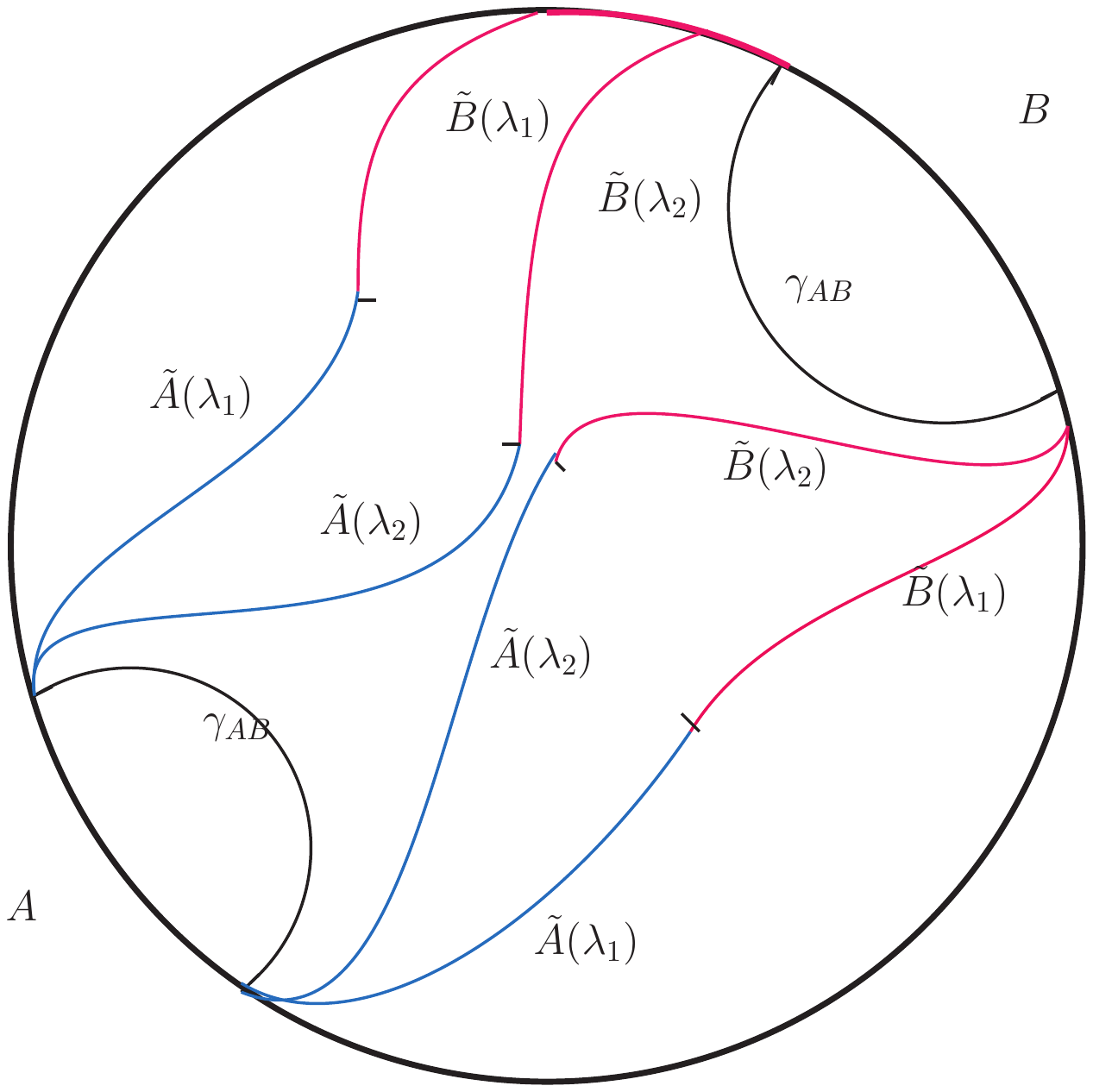}
\caption{A series of deformation of $\td A$ (blue) and $\td B$ (red) for the disconnected entanglement wedge. }
\label{f2}
\end{figure}

\begin{figure}[H]
\centering
\includegraphics[trim = 0mm 140mm -20mm 0mm, clip=true,width=10.0cm]{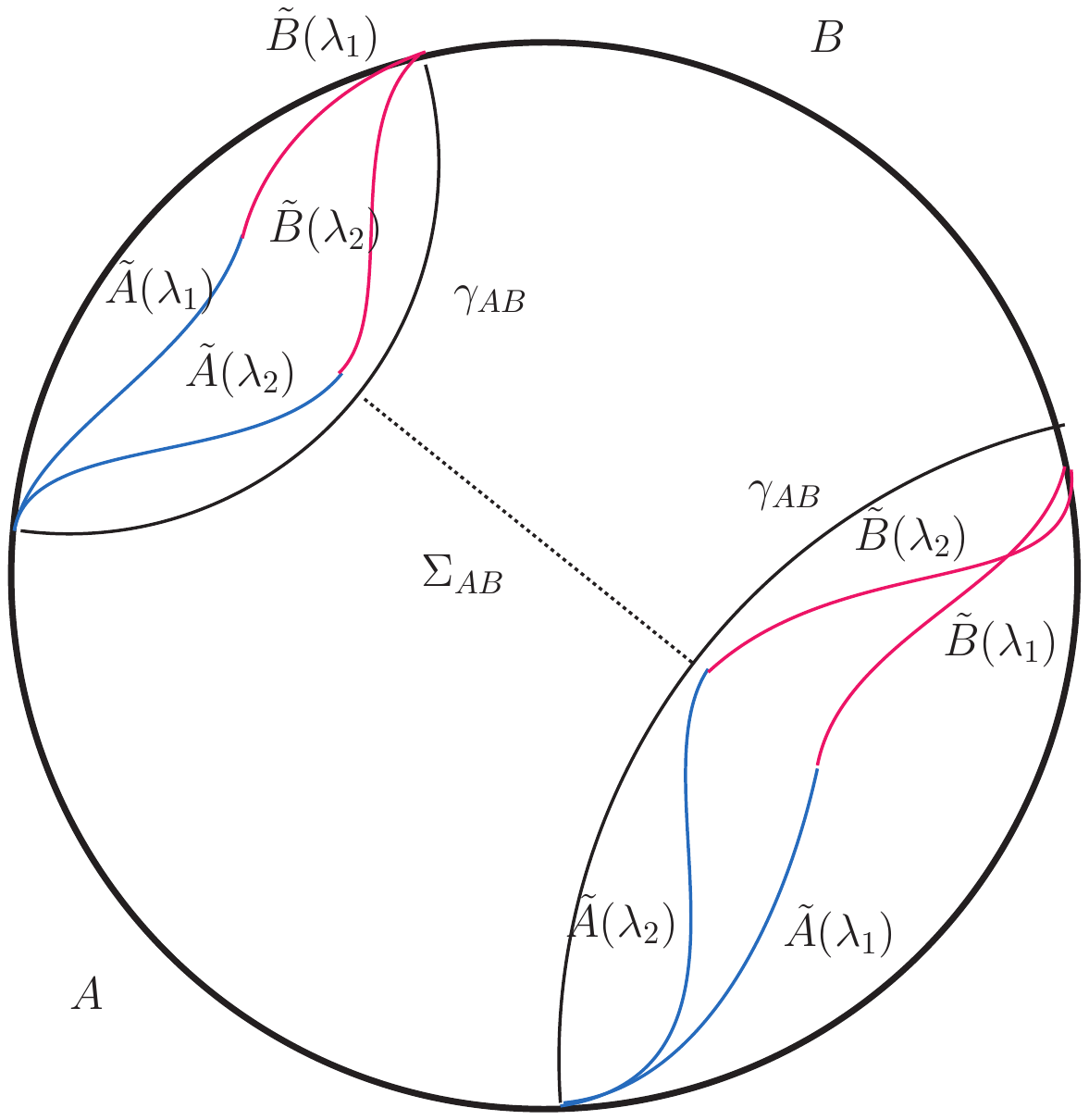}
\caption{A series of deformation of $\td A$ (blue) and $\td B$ (red) for the connected entanglement wedge. }
\label{f3}
\end{figure}

Even though we have constrained the set of the purifications to be $\mathcal{H}_\psi$, it is still hard to find the minimization of $S_{A\td A}$ by directly calculating in field theory. In the paper \cite{Miyaji:2015yva} the authors proposed a new duality relation between a bulk codimension-2 spacelike surface and quantum states in the dual field theory, which is expected to be a generalization of original AdS/CFT. In the context of surface/state correspondence, the gravity lives on a manifold $M_{d+2}$, any codimension-2 \textit{convext} surface $\Sigma$ corresponds to state in the total Hilbert space $\mathcal{H}$. In this paper we would work in AdS$_3$, the states are represented by curves $\sigma$ in AdS space. We would like to summarize the three important points of this correspondence:\\
1. A pure state $|\Phi(\sigma)\rangle$ corresponds to topologically trivial curve, i.e., homologous to a point, in the bulk.\\
2. If two curves $\sigma_1$ and $\sigma_2$ are connected by a smooth deformation that preserves \textit{convexity}, the corresponding states of them are related by
a unitary transformation, that is
\be
|\Phi(\sigma_1)\rangle=U(1,2)|\Phi(\sigma_2)\rangle,
\ee
where $U(1,2)$ is a unitary operator associated with deformation pathes.\\
3. The entanglement entropy for a subregion $\sigma_A$ of the curve $\sigma$ is conjectured to be given by the area formula,
\be\label{RTgeneral}
S_{\sigma,A}= \frac{\text{min}\{\text{area}(\gamma_{\sigma,A})\}}{4G},
\ee
where $G$ is the Newton constant.\\
If taking the curve $\sigma$ to be AdS boundary, these would be the AdS/CFT correspondence, specially the entanglement entropy is RT formula.
According to {\textbf{Corollary 1}}, we are interested in the unitary transformation $\mathcal{U}_{\overline{AB}}$ that act on subregion $\overline{AB}$.
In the bulk these transformations are dual to deformations of curve on the AdS boundary while keeping the boundary of $\overline{AB}$ invariant. Note that for a unitary operator $\mathcal{U}_A$ located in a subregion $A$ acting on the state $\Phi(\sigma)$, the corresponding deformation of surface $\sigma$ cannot transcend the extremal surface $\gamma_{\sigma,A}$. Only in this way one could keep the \textit{convexity} of the deformed curves, and this also guarantees the holographic entanglement entropy of subregion $A$ is invariant under unitary transformation $\mathcal{U}_A$\cite{Miyaji:2015yva}.\\
Now we are ready to prove the holographic EoP based on the surface/state correspondence. For simplicity we choose $A$ and $B$ to be two disconnected interval as shown in Fig.\ref{f2}. \\
If $A$ and $B$ are far away from each other, the entanglement wedge $W_{AB}$, defined by a region surrounded by $A$, $B$ and the minimal surface $\gamma_{A,B}$ homologous to them, would become disconnected, see Fig.\ref{f2} .
In this case we may choose a series of deformations of the curve $\td A(\lambda)$ and $\td B(\lambda)$. Since these deformations correspond to unitary transformations $\mathcal{U}_{\overline{AB}}$, they just need to keep the boundary of $\td A(\lambda)\td B(\lambda)$ invariant. As shown in the Fig.\ref{f2} we always could choose a series of deformations $\td A(\lambda_n)$ and $\td B(\lambda_n)$ such that  $\td A(\lambda_\infty)=\lim_{n\to \infty}\td A(\lambda_n)$ becomes connected in the bulk. Recall the definition of EoP (\ref{definitionPurification}), it is equal to the minimal value of entanglement entropy $S_{A\td A}$. The holographic entanglement entropy for $A\tilde A(\lambda_n)$ is given by (\ref{RTgeneral}). Therefore, we get $S_{A\td A(\lambda_\infty)}=0$\cite{Explain3}. This means the holographic EoP is zero. Note that in the Fig.\ref{f2} we only draw a special example for the deformations. In principle, there exist infinite ways to make $S_{A\td A(\lambda_\infty)}=0$. For example the deformations corresponding to $\mathcal{U}_{\td A}$ or $\mathcal{U}_{\td B}$ would never effect the value of $S_{A\td A(\lambda_n)}$.\\
If the entanglement wedge $W_{AB}$ becomes connected as shown in Fig.\ref{f3}, a series of deformations $\td A(\lambda)$ and $\td B(\lambda)$ corresponding to the unitary transformations $\mathcal{U}_{\overline{AB}}$ still keep the boundary of $\td A(\lambda)\td B(\lambda)$ invariant. One of the examples is shown in Fig.\ref{f3}. In this case the curve of $\td A(\lambda_n)$ would never possible become connected, since the deformation should never transcend the extremal surface $\gamma_{AB}$. Suppose $\Sigma_{AB}$ is the extremal surface as well as minimal area with the end points on the extremal surface $\gamma_{AB}$.\\
  Therefore, to get the minimal value of $S_{A\td A}$ one could construct a series of deformations such that the end points of $\td A(\lambda_\infty)=\lim_{n\to \infty}\td A(\lambda_n)$ coincide with the ones of $\Sigma_{AB}$. In this limit we would have the minimal value of $S_{A\td A}$ which is given by
\be
S_{ A\td A(\lambda_\infty)}=\frac{\Sigma_{AB}}{4G}.
\ee
Therefore, we have proved the holographic EoP in the context of surface/state correspondence. In above discussion we only focus on two intervals case, but it is straightforward to generalize the proof to more complicated cases. The holographic generalization to muti-partite correlations is discussed in \cite{Umemoto:2018jpc}.

\section{Projective Measurement and EoP in CFT}
\begin{figure}[H]
\centering
\includegraphics[trim = 0mm 170mm -40mm 0mm, clip=true,width=10.0cm]{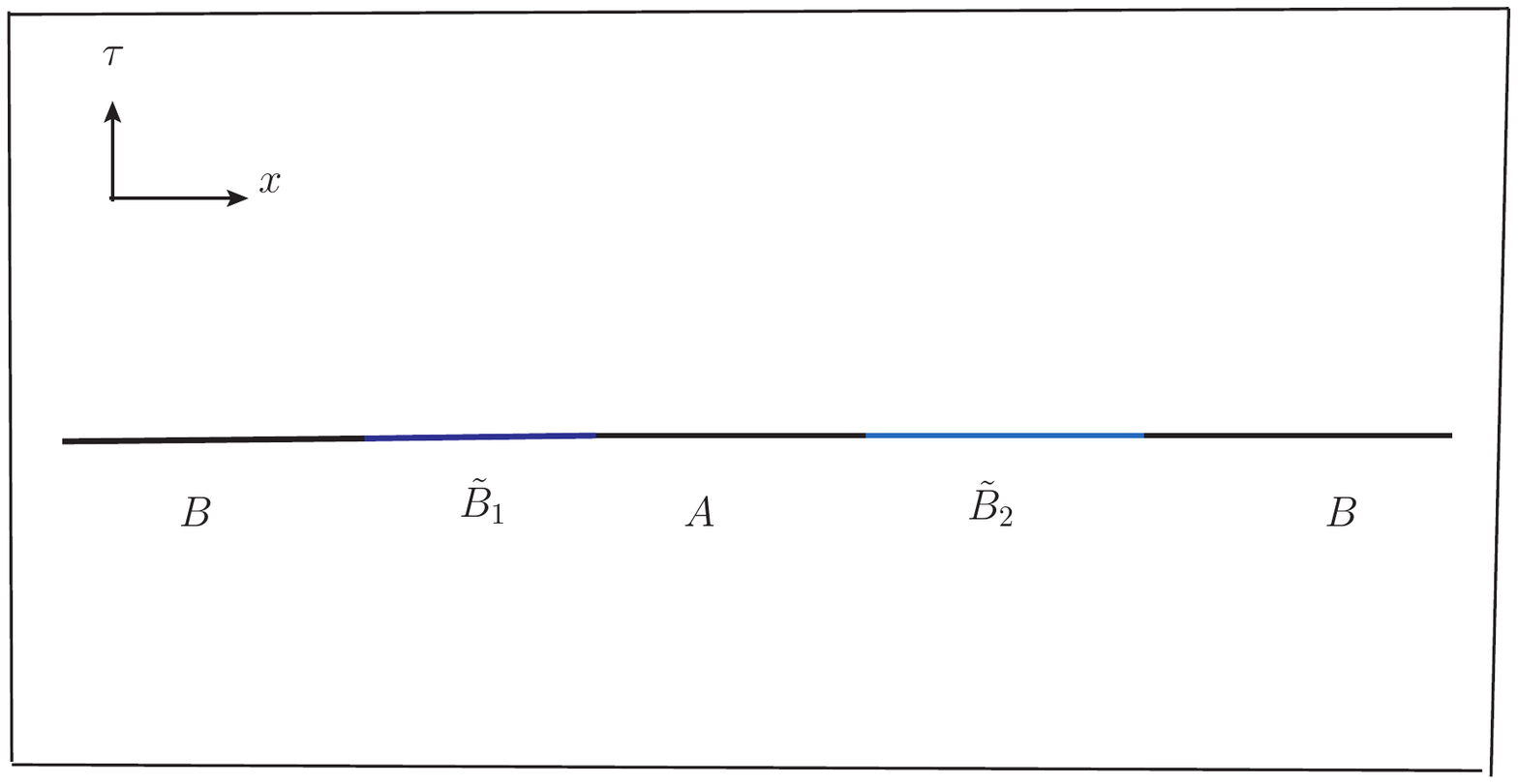}
\caption{The state with projective measurement in region $\td B:=\td B_1 \td B_2$ (blue line) can be represented by path integral on the lower half plane with two slits.}
\label{f4}
\end{figure}

Another interesting question is what kinds of unitary operators $\mathcal{U}_{\overline{AB}}$ would produce the minimal value of $S_{A\td A}$. Firstly, we should note that the unitary operator is not unique, if $\mathcal{\td U}_{\overline{AB}}$ is one, so are the operators $\mathcal{U}_{\bar A}\mathcal{\td U}_{\overline{AB}}$ and $\mathcal{U}_{\bar B}\mathcal{\td U}_{\overline{AB}}$. It is still an open question whether the operator is unique up to the above gauge.\\
In this section we will discuss one special operator belonging to the algebra $\mathscr{R}(\overline{AB})$, that is the projective measurement in CFT. The projective measurement in 1+1 dimensional CFT was discussed in \cite{Rajabpour:2015uqa}\cite{Rajabpour:2015xkj}\cite{Najafi:2016kwb}, its holographic explanation and applications can be found in \cite{Numasawa:2016emc}. We focus on the projective measurement $\mathcal{P}^{\alpha}_{\overline{AB}}$, which makes the states in the region $\overline{AB}$ fixed by some conformal bases $\alpha$. For example, for free boson theory, a projective measurement with fixed $\phi$ in region $A$ corresponds to Dirichlet boundary condition on $A$, which is a conformal boundary.\\
Note that the projective measurements $\mathcal{P}^{\alpha}_{\overline{AB}}$ are not unitary. But as we will show soon the entanglement entropy $S_{A\td A}$ with $\td B= \overline{AB}$ is very close to the holographic EoP.\\
We would follow the results in \cite{Najafi:2016kwb}, there the author considered the projective measurement is for two intervals as shown in Fig.\ref{f4}. The projective state $\mathcal{P}_{\td B}|0\rangle$ can be represented by path integral on the lower half plane with two slits on $\td{B}_1$ and $\td B_2$. Assume the length of the intervals $l_{\td B_1}=s_1$, $l_{\td B_2}=s_2$ and $l_A=l$. To calculate R\'enyi entropy $S^n_{A}$ for subsystem $A$ in the state $\mathcal{P}_{\td B}|0\rangle$ we need to evaluate the path integral on the n-sheet surface $\Sigma_n$ with two slits $\td B_1$,$\td B_2$ and branch cut on $A$. The R\'enyi entropy is given by
\be
S_A^n=\frac{1}{1-n}\log \frac{\mathcal{Z}_{\Sigma_n}}{\mathcal{Z}_{\Sigma_1}^n},
\ee
where $\Sigma_1$ is the surface with two slits. The entanglement entropy is just $S_A=\lim_{n\to 1}S^n_A$. The partition function $\mathcal{Z}_{\Sigma_n}$ can be calculated throw a conformal mapping $w_n(z)$ from $\Sigma_n$ to annulus, see Appendix of \cite{Rajabpour:2015xkj} for the detail of the mapping.
For general $s_1,s_2, l$ there are no analytical results . In the limit $l\gg s_1=s_2=s$ the result is
\be\label{limit1}
S_A=\frac{c}{3} \log \frac{l}{s}+...,
\ee
where $...$ denote the contributions from the boundary, which are not related to central charge $c$.
In the limit $l\ll s_1=s_2=s$, $S_A=0$ up to some boundary contributions.\\
In the limit $s_2\ll s_1,l$,
\be\label{limit2}
S_A=\frac{c}{6}\log \frac{l(l+s_1)}{s_1s_2}+...,
\ee
with $...$ being the boundary contributions. \\
Now we would like to compare the entanglement entropy after projective measurement with holographic EoP for $AB$. In the limit $l\ll s_1=s_2=s$, the entanglement wedge of $AB$ becomes disconnected, the holographic EoP is vanishing. The entanglement entropy after projective measurement is also vanishing up to some boundary contributions.\\
In the limit $l\gg s_1=s_2=s$ or $s_2\ll s_1,l$ the entanglement wedge of $AB$ should be connected.
In the appendix we calculate the holographic EoP for the interval $A$ and $B$, the result is
\bea
&&E_{AB}=\frac{c}{6}\log\Big[\frac{s_1 \left(2 l+s_2\right)}{s_1 s_2}\nn \\
&&+\frac{2 \left(l^2+l s_2+\sqrt{l \left(l+s_1\right) \left(l+s_2\right) \left(l+s_1+s_2\right)}\right)}{s_1 s_2}\Big].
\eea
The holographic EoP is
\begin{equation}
E_{AB}=\left\{
\begin{aligned}
&&\frac{c}{3}\log \frac{2 l}{s},\ \text{in the limit}\  l\gg s_1=s_2=s\\ \nn
&&\frac{c}{6}\log \frac{4 l \left(l+s_1\right)}{s_1 s_2},\ \text{in the limit}\  s_2\ll s_1,l.
\end{aligned}
\right.
\end{equation}
Comparing with the results of projective measurement (\ref{limit1}) and (\ref{limit2}) in the same limit, we find that their difference is $\frac{c}{3}\log 2$. This suggests the projective measurement operator may provide as an approximate operator of the unitary operator that produces the minimal value of $S_{A\td A}$.
\section{Discussions}
Our discussions are mainly for vacuum state, but it is straightforward to generalize to other cyclic state, such as the states on which the translation group acts homorphically\cite{Witten:2018lha}. For mixed state in 1+1 dimension CFT the thermal state is conformal equal to the vacuum state, our discussions may be generalized to that case. This may fail for non-entangled state, such as the boundary state in CFT\cite{Cardy}\cite{Miyaji:2014mca}.\\
Our proof of holographic EoP only includes the states that can be described by geometry in the bulk. At least in 2D CFT it is expected there are many states that cannot be dual to a classical geometry\cite{Guo:2018fnv}. So the proof is only true for the class of geometric states.\\
The small difference between holographic EoP and entanglement entropy after projective measurement may be understood along with holographic explanation of projective measurement\cite{Numasawa:2016emc}. We would explore more on this in the near future.\\
~\\
\quad \textit{I would like to thank Pak Hang Chris Lau for discussions.  I also would like to thank the organisers of the NCTS Annual Meeting 2018: Particles, Cosmology and String. Tadashi Takayanagi's talk at this conference brought my attention to entanglement of purification.  I am  supported in part by the National Center of Theoretical Science (NCTS). }

\appendix
\section{Holographic EoP of two intervals in 1+1 dimensional CFT}
\begin{figure}[H]
\centering
\includegraphics[trim = 0mm 170mm -40mm 0mm, clip=true,width=12.0cm]{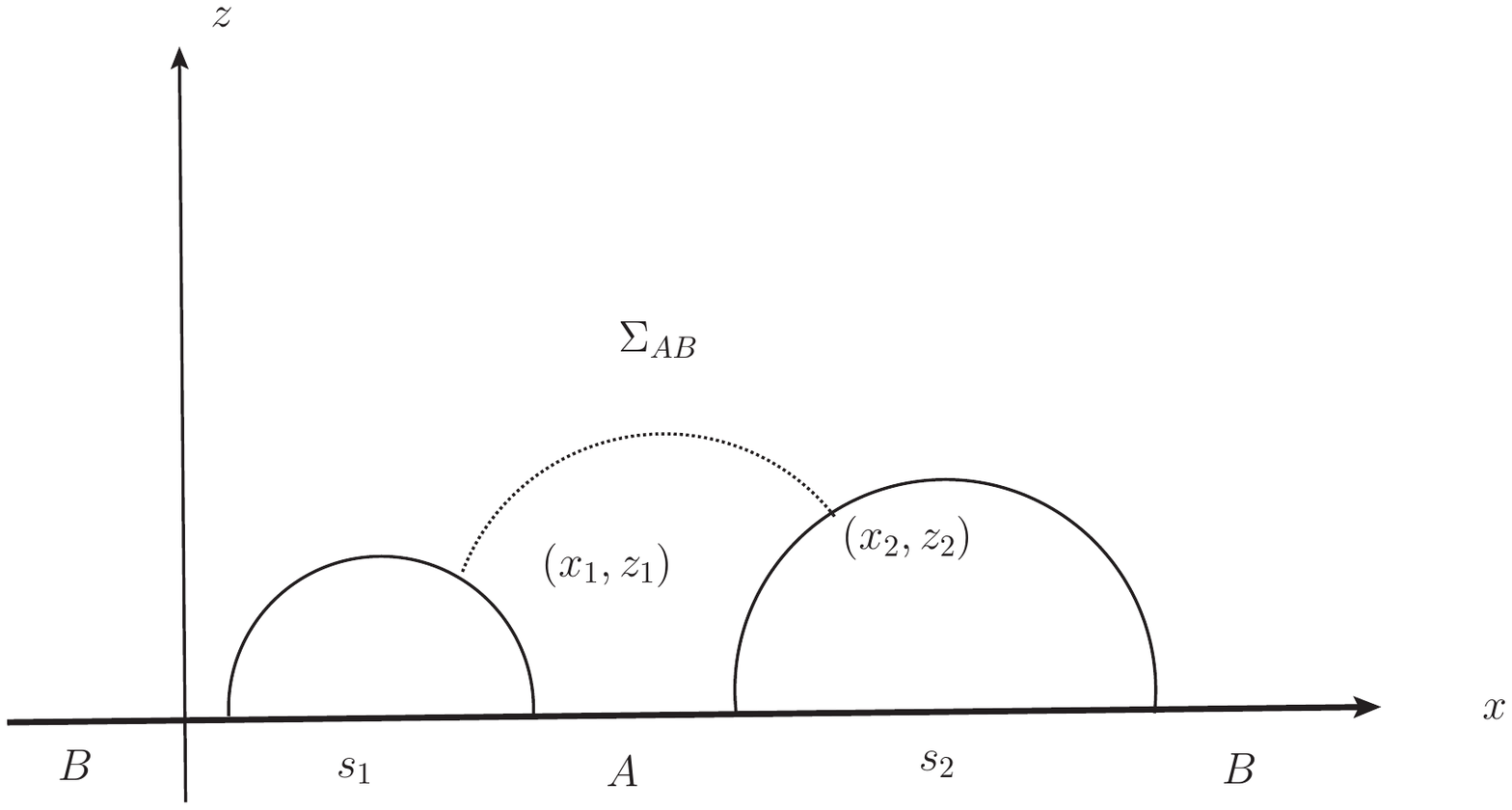}
\caption{Calculations of holographic EoP for two interval $A$ and $B$ with $l_A=l$ and the distances between $A$, $B$ are $s_1$ and $s_2$. }
\label{f5}
\end{figure}

We will derive the holographic EoP of two intervals in 1+1 dimensional CFT in this section. To compare with the result in the main tex we choose  $A$ and $B$ as shown in Fig.\ref{f4}. We only plot the case when $AB$ has connected entanglement wedge in Fig.\ref{f5}. To calculate holographic EoP we need to compute the length of the entanglement wedge cross section, i.e., $\Sigma_{AB}$ in Fig.\ref{f5}. The minimal length condition leads to the curve $\Sigma_{AB}$ is perpendicular to the extremal surface of entanglement wedge at the points $(x_1,z_1)$ and $(x_2,z_2)$. With some simple calculations we get
\bea
&&z_1=\frac{s_1 \sqrt{l \left(l+s_1\right) \left(l+s_2\right) \left(l+s_1+s_2\right)}}{s_1^2+2 l \left(l+s_2\right)+s_1 \left(2 l+s_2\right)}\nn\\
&&z_2=\frac{s_2 \sqrt{l \left(l+s_1\right) \left(l+s_2\right) \left(l+s_1+s_2\right)}}{2 l^2+2 l \left(s_1+s_2\right)+s_2 \left(s_1+s_2\right)},
\eea
and the equation of the curve $\Sigma_{AB}$, $(x-x_0)^2+z^2=z_*^2$ with
\bea
&&x_0=\frac{l(s_2-s_2)}{2(2l+s_1+s_2)}\nn\\
&&z_*=\frac{\sqrt{l \left(l+s_1\right) \left(l+s_2\right) \left(l+s_1+s_2\right)}}{2 l+s_1+s_2}.
\eea
The length of $\Sigma_{AB}$ is
\bea
&&L=\log\Big[\frac{s_1 \left(2 l+s_2\right)}{s_1 s_2}\nn \\
&&+\frac{2 \left(l^2+l s_2+\sqrt{l \left(l+s_1\right) \left(l+s_2\right) \left(l+s_1+s_2\right)}\right)}{s_1 s_2}\Big].
\eea

\begin{thebibliography}{00}
%\cite{Maldacena:1997re}
\bibitem{Maldacena:1997re}
  J.~M.~Maldacena,
  %``The Large N limit of superconformal field theories and supergravity,''
  Int.\ J.\ Theor.\ Phys.\  {\bf 38}, 1113 (1999)
  [Adv.\ Theor.\ Math.\ Phys.\  {\bf 2}, 231 (1998)]
  %doi:10.1023/A:1026654312961, 10.4310/ATMP.1998.v2.n2.a1
  [hep-th/9711200].
%\cite{Ryu:2006bv}
\bibitem{Ryu:2006bv}
  S.~Ryu and T.~Takayanagi,
  %``Holographic derivation of entanglement entropy from AdS/CFT,''
  Phys.\ Rev.\ Lett.\  {\bf 96}, 181602 (2006)
  %doi:10.1103/PhysRevLett.96.181602
  [hep-th/0603001].
  %\cite{Hubeny:2007xt}
\bibitem{Hubeny:2007xt}
  V.~E.~Hubeny, M.~Rangamani and T.~Takayanagi,
  %``A Covariant holographic entanglement entropy proposal,''
  JHEP {\bf 0707}, 062 (2007)
  %doi:10.1088/1126-6708/2007/07/062
  [arXiv:0705.0016 [hep-th]].
%\cite{Miyaji:2015yva}
\bibitem{Haag}
R. Haag, Local quantum physics: Fields, particles, algebras. Springer Science  Business Media, 2012.
\bibitem{Streater}
R. F. Streater and  A. S.  Wightman,  PCT, spin and statistics, and all that. Princeton University Press,2016.
\bibitem{Terhal}
B. M. Terhal, M. Horodecki, D. W. Leung and D. P.  DiVincenzo,
%The entanglement of purification.
Journal of Mathematical Physics, 43(9), 4286-4298 (2002).
[arXiv:0202044[quant-ph]]

%\cite{Takayanagi:2017knl}
\bibitem{Takayanagi:2017knl}
  T.~Takayanagi and K.~Umemoto,
  %``Entanglement of purification through holographic duality,''
  Nature Phys.\  {\bf 14}, no. 6, 573 (2018)
  %doi:10.1038/s41567-018-0075-2
  [arXiv:1708.09393 [hep-th]].
%\cite{Nguyen:2017yqw}
\bibitem{Nguyen:2017yqw}
  P.~Nguyen, T.~Devakul, M.~G.~Halbasch, M.~P.~Zaletel and B.~Swingle,
  %``Entanglement of purification: from spin chains to holography,''
  JHEP {\bf 1801}, 098 (2018)
 % doi:10.1007/JHEP01(2018)098
  [arXiv:1709.07424 [hep-th]].
%\cite{Czech:2012bh}
\bibitem{Czech:2012bh}
  B.~Czech, J.~L.~Karczmarek, F.~Nogueira and M.~Van Raamsdonk,
  %``The Gravity Dual of a Density Matrix,''
  Class.\ Quant.\ Grav.\  {\bf 29}, 155009 (2012)
  %doi:10.1088/0264-9381/29/15/155009
  [arXiv:1204.1330 [hep-th]].
  %\cite{Wall:2012uf}
\bibitem{Wall:2012uf}
  A.~C.~Wall,
  %``Maximin Surfaces, and the Strong Subadditivity of the Covariant Holographic Entanglement Entropy,''
  Class.\ Quant.\ Grav.\  {\bf 31}, no. 22, 225007 (2014)
  %doi:10.1088/0264-9381/31/22/225007
  [arXiv:1211.3494 [hep-th]].
%\cite{Headrick:2014cta}
\bibitem{Headrick:2014cta}
  M.~Headrick, V.~E.~Hubeny, A.~Lawrence and M.~Rangamani,
  %``Causality & holographic entanglement entropy,''
  JHEP {\bf 1412}, 162 (2014)
  %doi:10.1007/JHEP12(2014)162
  [arXiv:1408.6300 [hep-th]].
%\cite{Dong:2016eik}
\bibitem{Dong:2016eik}
  X.~Dong, D.~Harlow and A.~C.~Wall,
  %``Reconstruction of Bulk Operators within the Entanglement Wedge in Gauge-Gravity Duality,''
  Phys.\ Rev.\ Lett.\  {\bf 117}, no. 2, 021601 (2016)
  %doi:10.1103/PhysRevLett.117.021601
  [arXiv:1601.05416 [hep-th]].
  %\cite{Caputa:2018xuf}
\bibitem{Caputa:2018xuf}
  P.~Caputa, M.~Miyaji, T.~Takayanagi and K.~Umemoto,
  %``Holographic Entanglement of Purification from Conformal Field Theories,''
  arXiv:1812.05268 [hep-th].
\bibitem{Hauschild}
J. Hauschild, Johannes, E. Leviatan; J. Bardarson,
E. Altman, M. Zaletel, F. Pollmann, “Finding
purifications with minimal entanglement ,” eprint
arXiv:1711.01288
%\cite{Bhattacharyya:2018sbw}
\bibitem{Bhattacharyya:2018sbw}
  A.~Bhattacharyya, T.~Takayanagi and K.~Umemoto,
  %``Entanglement of Purification in Free Scalar Field Theories,''
  JHEP {\bf 1804}, 132 (2018)
  %doi:10.1007/JHEP04(2018)132
  [arXiv:1802.09545 [hep-th]].
\bibitem{Miyaji:2015yva}
  M.~Miyaji and T.~Takayanagi,
  %``Surface/State Correspondence as a Generalized Holography,''
  PTEP {\bf 2015}, no. 7, 073B03 (2015)
 % doi:10.1093/ptep/ptv089
  [arXiv:1503.03542 [hep-th]].
  %\cite{Miyaji:2015fia}
\bibitem{Miyaji:2015fia}
  M.~Miyaji, T.~Numasawa, N.~Shiba, T.~Takayanagi and K.~Watanabe,
  %``Continuous Multiscale Entanglement Renormalization Ansatz as Holographic Surface-State Correspondence,''
  Phys.\ Rev.\ Lett.\  {\bf 115}, no. 17, 171602 (2015)
  %doi:10.1103/PhysRevLett.115.171602
  [arXiv:1506.01353 [hep-th]].
 %\cite{Rajabpour:2015uqa}
\bibitem{Rajabpour:2015uqa}
  M.~A.~Rajabpour,
  %``Post measurement bipartite entanglement entropy in conformal field theories,''
  Phys.\ Rev.\ B {\bf 92}, no. 7, 075108 (2015)
  %doi:10.1103/PhysRevB.92.075108
  [arXiv:1501.07831 [cond-mat.stat-mech]].
  %\cite{Rajabpour:2015xkj}
\bibitem{Rajabpour:2015xkj}
  M.~A.~Rajabpour,
  %``Entanglement entropy after a partial projective measurement in $1+1$ dimensional conformal field theories: exact results,''
  J.\ Stat.\ Mech.\  {\bf 1606}, no. 6, 063109 (2016)
  %doi:10.1088/1742-5468/2016/06/063109
  [arXiv:1512.03940 [hep-th]].
  \bibitem{Kay}
B. S. Kay and R. M. Wald,
%Theorems on the uniqueness and thermal properties of stationary, nonsingular, quasifree states on spacetimes with a bifurcate Killing horizon.
Physics Reports, 207(2), 49-136 (1991).
%\cite{Witten:2018lha}
\bibitem{Witten:2018lha}
  E.~Witten,
  %``APS Medal for Exceptional Achievement in Research: Invited article on entanglement properties of quantum field theory,''
  Rev.\ Mod.\ Phys.\  {\bf 90}, no. 4, 045003 (2018)
  %doi:10.1103/RevModPhys.90.045003
  [arXiv:1803.04993 [hep-th]].

\bibitem{Explain1}
More precisely, for any state $|\psi\rangle$ one always could find an operator $\mathcal{O}_{\overline{AB}}(\psi)$ such that the $||\psi\rangle- \mathcal{O}_{\overline{AB}}(\psi)|0\rangle|< \epsilon$, where $\epsilon$ is an aribrary positive constant.
\bibitem{Explain2}
In fact this follows by another important property of vacuum state, that is the separating property. A state $|\Omega\rangle$ is said to be separating for the local algebra $\mathscr{R}(O)$ if $\mathcal{O}|\Omega\rangle =0\Rightarrow \mathcal{O}=0$.

 \bibitem{Explain3}
 The original state corresponding to $A\td A$ is a mixed state, since it is a part of a closed  surface, i.e., AdS boundary. Under a series of unitary transformations, the state would approach to a pure state. This is consistent with the intuition of purification process.
 %\cite{Umemoto:2018jpc}
\bibitem{Umemoto:2018jpc}
  K.~Umemoto and Y.~Zhou,
  %``Entanglement of Purification for Multipartite States and its Holographic Dual,''
  JHEP {\bf 1810}, 152 (2018)
  %doi:10.1007/JHEP10(2018)152
  [arXiv:1805.02625 [hep-th]].
  %\cite{Najafi:2016kwb}
\bibitem{Najafi:2016kwb}
  K.~Najafi and M.~A.~Rajabpour,
  %``Entanglement entropy after selective measurements in quantum chains,''
  JHEP {\bf 1612}, 124 (2016)
  %doi:10.1007/JHEP12(2016)124
  [arXiv:1608.04074 [cond-mat.str-el]].
  %\cite{Numasawa:2016emc}
\bibitem{Numasawa:2016emc}
  T.~Numasawa, N.~Shiba, T.~Takayanagi and K.~Watanabe,
  %``EPR Pairs, Local Projections and Quantum Teleportation in Holography,''
  JHEP {\bf 1608}, 077 (2016)
 % doi:10.1007/JHEP08(2016)077
  [arXiv:1604.01772 [hep-th]].
  \bibitem{Cardy}
  J. L. Cardy, “Boundary Conditions, Fusion Rules and the Verlinde Formula,” Nucl.
Phys. B 324 (1989) 581.

  %\cite{Miyaji:2014mca}
\bibitem{Miyaji:2014mca}
  M.~Miyaji, S.~Ryu, T.~Takayanagi and X.~Wen,
  %``Boundary States as Holographic Duals of Trivial Spacetimes,''
  JHEP {\bf 1505}, 152 (2015)
  %doi:10.1007/JHEP05(2015)152
  [arXiv:1412.6226 [hep-th]].
  %\cite{Guo:2018fnv}
\bibitem{Guo:2018fnv}
  W.~Z.~Guo, F.~L.~Lin and J.~Zhang,
  %``Non-geometric States in a Holographic Conformal Field Theory,''
  arXiv:1806.07595 [hep-th].
\end{thebibliography}
\end{document}